\documentclass[sigconf]{acmart}
\AtBeginDocument{%
  }

\setcopyright{acmlicensed}
\copyrightyear{2018}
\acmYear{2018}
\acmDOI{XXXXXXX.XXXXXXX}

\acmConference[Conference acronym 'XX]{Make sure to enter the correct
  conference title from your rights confirmation emai}{June 03--05,
  2018}{Woodstock, NY}
\acmISBN{978-1-4503-XXXX-X/18/06}

\usepackage{algorithm}
\usepackage{algpseudocode}
\usepackage{siunitx}
\usepackage{anyfontsize}
\usepackage{graphicx} 
\usepackage{amsmath}
\usepackage{mathrsfs} 
\usepackage{pict2e}
\usepackage{graphicx}
\usepackage{subcaption}

\newtheorem{definition}{Definition}[section]
\newtheorem*{theorem*}{\bf Research Problem}
\usepackage{comment}
\usepackage{multirow}
\usepackage{multicol}
\usepackage{makecell}
\usepackage{enumitem}
\usepackage{graphicx}
\usepackage{subcaption}
\begin{document}

\title{TrajRoute: Rethinking Routing with a Simple Trajectory-Based Approach — Forget the Maps and Traffic!}

\author{Maria Despoina Siampou}
\email{siampou@usc.edu}
\affiliation{%
  \institution{University of Southern California}
  \city{Los Angeles}
  \state{California}
  \country{USA}
}

\author{Chrysovalantis Anastasiou}
\email{canastas@usc.edu}
\affiliation{%
  \institution{University of Southern California}
  \city{Los Angeles}
  \state{California}
  \country{USA}
}

\author{John Krumm}
\email{jkrumm@usc.edu}
\affiliation{%
  \institution{University of Southern California}
  \city{Los Angeles}
  \state{California}
  \country{USA}
}

\author{Cyrus Shahabi}
\email{shahabi@usc.edu}
\affiliation{%
  \institution{University of Southern California}
  \city{Los Angeles}
  \state{California}
  \country{USA}
}

\renewcommand{\shortauthors}{Siampou et al.}

\newcommand{\modelname}{\textsf}


\begin{abstract}
The abundance of vehicle trajectory data offers a new opportunity to compute driving routes between origins and destinations. Current graph-based routing pipelines, while effective, involve substantial costs in constructing, maintaining, and updating road network graphs to reflect real-time conditions. In this study, we propose a new trajectory-based routing paradigm that bypasses current workflows by directly utilizing raw trajectory data to compute efficient routes. Our method, named \modelname{TrajRoute}, uniquely ``follows'' historical trajectories from a source to a destination, constructing paths that reflect actual driver behavior and implicit preferences. To supplement areas with sparse trajectory data, the road network is also incorporated into \modelname{TrajRoute}'s index, and tunable parameters are introduced to control the balance between road segments and trajectories, ensuring a unified and adaptable routing approach. We experimentally verify our approach by comparing it to an existing online routing service. Our results demonstrate that as the number of trajectories covering the road network increases, \modelname{TrajRoute} produces increasingly accurate travel time and route length estimates while gradually eliminating the need to downgrade to the road network. This highlights the potential of simpler, data-driven pipelines for routing, offering lower-maintenance alternatives to conventional systems.  

\end{abstract}



\received{17 October 2024}

\maketitle

\section{Introduction}

\begin{figure*}[ht]
    \centering
    \begin{subfigure}[b]{0.9\linewidth}
        \centering
        \includegraphics[width=\linewidth]{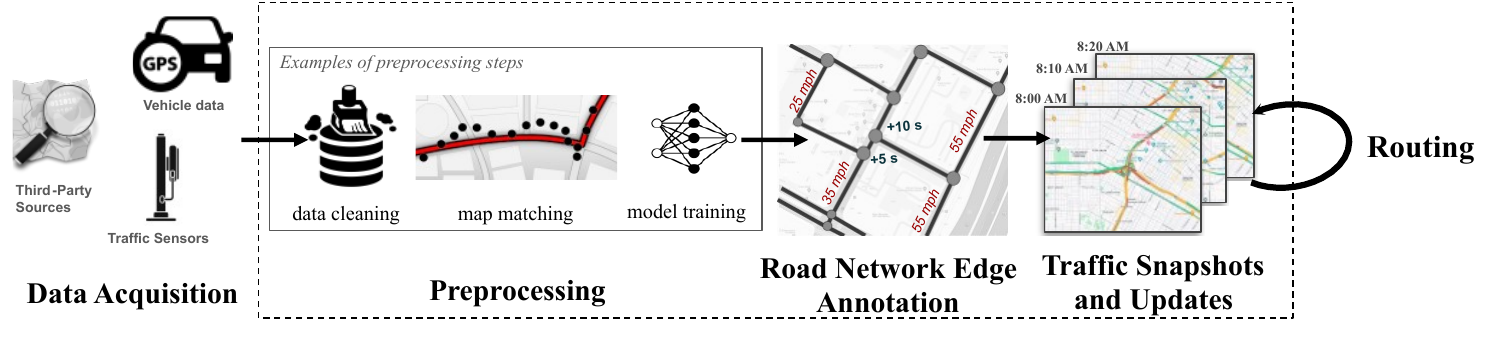}
        \caption{Overview of a typical pipeline used by current routing services. The steps in the dashed box are repeated as newer data becomes available.}
        \label{fig:conv-pipeline}
    \end{subfigure}
    
    \vspace{0.5em} 
    
    \begin{subfigure}[b]{0.7\linewidth}
        \centering
        \includegraphics[width=\linewidth]{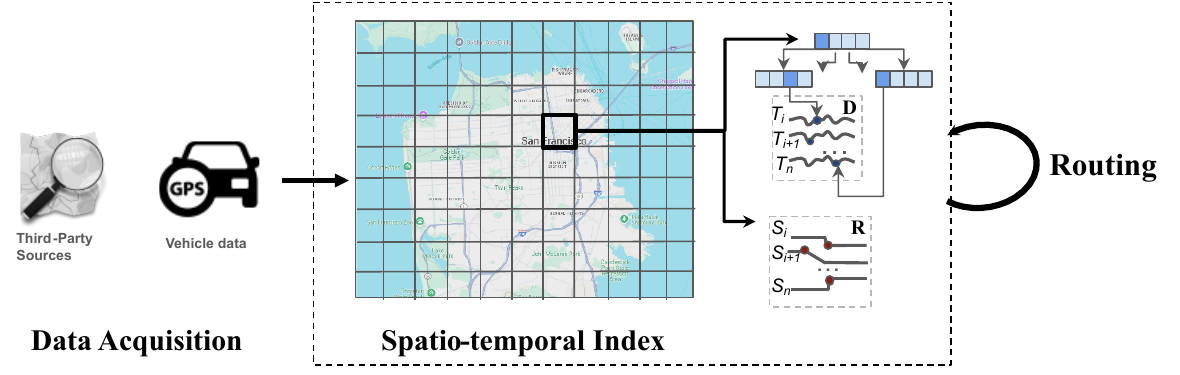}
        \caption{Our proposed, \modelname{TrajRoute} pipeline.}
        \label{fig:our-pipeline}
    \end{subfigure}
    
    \caption{Comparison of pipelines: (a) current routing services and (b) our proposed pipeline.}
    \label{fig:pipelines-overview}
    \Description{Comparison of pipelines: (a) current routing services and (b) our proposed pipeline.}
\end{figure*}

The recent widespread adoption of location-acquisition technologies, such as GPS data from vehicles, has generated vast amounts of user trajectory data~\cite{ding2015enabling, hu2018risk}. For instance, companies such as Tesla and Google have access to comprehensive real-time trajectory datasets that their users continuously provide. This has offered a foundation for improving navigation services, where trajectory data are used to annotate the road network edges with more realistic travel time estimates~\cite{dai2016path}. As more data becomes available, these edges are continuously updated to reflect real-time conditions. A typical pipeline used by routing services, shown in Figure~\ref{fig:conv-pipeline}, involves several cost-intensive preprocessing steps. One common step is map matching, where GPS data is aligned with the road network~\cite{newson2009hidden}. The network is then dynamically updated with this information, generating traffic snapshots that enable accurate path recommendations. In an effort to deliver personalized routing, a number of research studies have also focused on utilizing these trajectory data to extract more insights such as preferences, patterns, and popular paths which are also used to annotate the graph edges, allowing for more customized route recommendations~\cite{ceikute2015vehicle, guo2018learning, chen2011discovering, delling2015navigation, letchner2006trip, patel2006personalizing, yang2015toward, chang2011discovering}. However, maintaining these graph-based pipelines is costly due to the need for continuous data collection, processing, and frequent updates to reflect real-time traffic estimates, rendering the entire process resource-intensive. 

\begin{figure}
    \centering
    \includegraphics[width=0.75\linewidth]{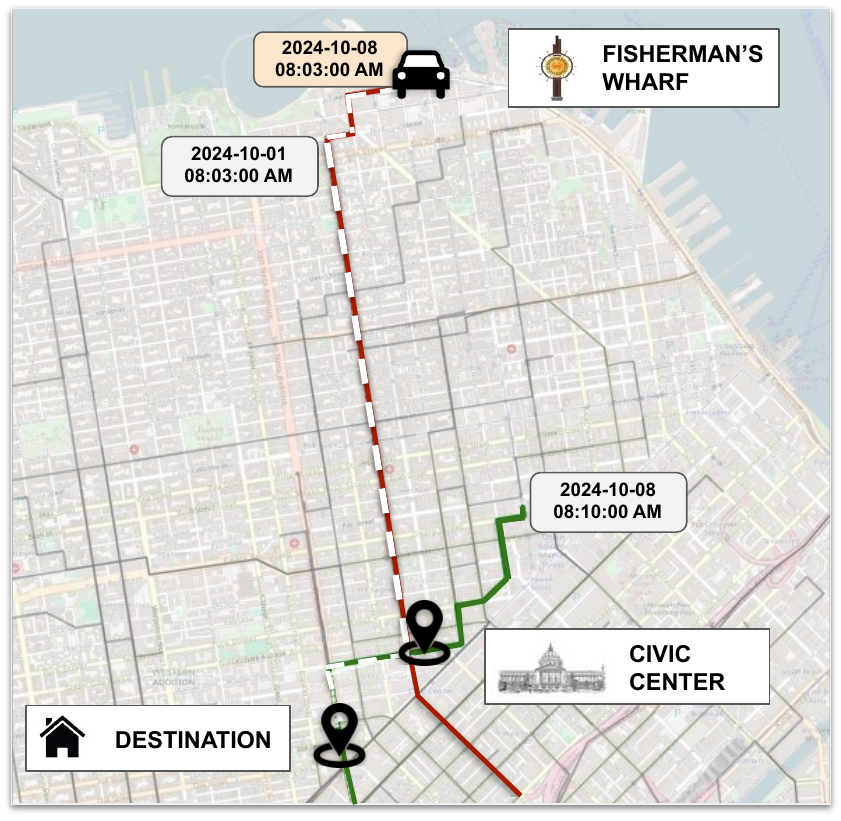}
        \caption{Example of \modelname{TrajRoute} generating a route from Fisherman's Wharf to the Mission District.
        }
    \label{fig:example}
    \Description{Example of \modelname{TrajRoute} generating a route from Fisherman's Wharf to the Mission District.}
\end{figure}

With the increasing availability of large-scale and up-to-date trajectory datasets, we have identified a new opportunity: the ability to extract routes directly from historical raw data. This approach eliminates the need for costly graph construction and ongoing maintenance, streamlining the process and leveraging the richness of real-world data. To this end, we propose a new approach for routing, namely \modelname{TrajRoute}, that directly leverages raw trajectory data for routing decisions. In this paper, we demonstrate that accurate route recommendations can be derived by following historical trajectories. \modelname{TrajRoute}'s approach is straightforward yet effective: given an origin, a destination, and a start time, the system utilizes historical trajectories to compute the optimal route in terms of travel time. If a complete historical trajectory between the origin and destination is unavailable, \modelname{TrajRoute} combines segments from multiple trajectories to construct a continuous path. Consider it analogous to hopping from one vehicle to another to reach the final destination. The following example, shown in Figure~\ref{fig:example}, illustrates how \modelname{TrajRoute} operates in practice.

{\it Example.} A user starts at Fisherman's Wharf in San Francisco, CA, and requests a route to their home near the Mission District using \modelname{TrajRoute} at 8:00 AM on a weekday. The system first identifies a commuter’s trajectory, marked in red, that was recorded around the same time and day in the past, which the user follows from Fisherman's Wharf to the Civic Center (shown with dashed white lines). Based on historical data, the system calculates that the user will arrive at the Civic Center at 8:20 AM. At this point, the system selects a second trajectory, marked in green, that departed at 8:10 AM the same day from Civic Center. This green trajectory leads to the Mission District, with the remainder of the journey estimated to take about 7 minutes, bringing the user to their destination at around 8:27 AM. While there are other historical routes recorded at the same time frame, denoted in grey color, this combination, from parts of two other trajectories, offers the fastest route based on historical information, ensuring an optimized and realistic travel experience for the user, while eliminating the need for the system to maintain or artificially determine road speeds and turn costs.

One key advantage of \modelname{TrajRoute} is that it eliminates the need for the costly preprocessing steps found in typical routing pipelines. As shown in Figure~\ref{fig:conv-pipeline} modern routing services rely on extensive data preprocessing, which includes tasks like map matching and continuously updating road network annotations as new data becomes available. The steps within the dashed box are reprocessed regularly to account for current traffic conditions. In contrast, \modelname{TrajRoute} bypasses these steps by working directly with raw historical trajectory data. To retrieve relevant trajectories efficiently, \modelname{TrajRoute} uses a spatio-temporal grid-based index, allowing it to find trajectories that best match a user's query. The approach performs optimally when there is consistently a segment of a trajectory near the current position and within the present time frame. For scenarios where a direct path to the destination is not entirely available from historical data, \modelname{TrajRoute} integrates adjacent road segments to bridge the gaps, ensuring that users will always receive a valid route as a query answer. 
It’s important to note that if no trajectories pass through these roads, conventional routing methods won’t have access to real-time travel times for those segments either. Figure~\ref{fig:our-pipeline}, presents the proposed framework.

We experimentally verify \modelname{TrajRoute} on a real trajectory dataset and compare the routes generated with those from Azure Maps\footnote{\url{https://azure.microsoft.com/en-us/products/azure-maps}} in terms of route accuracy and travel time precision. The results confirm that \modelname{TrajRoute} consistently delivers reliable and precise routing recommendations. One of the primary advantages of \modelname{TrajRoute} is its ability to inherently integrate routing features that are implicitly encoded in the trajectories, such as variable traffic speeds, intersection-related delays, and individual driver preferences, which are typically challenging and expensive to incorporate in standard graph-based routing systems. When supplied with abundant up-to-date trajectory data, \modelname{TrajRoute} can effectively eliminate the need for the heavy-weight pipeline of map-matching, traffic inference, and graph edge updating. Additionally, it is simple in tuning and maintenance, requiring only three parameters: the grid size, which determines the granularity of trajectory indexing, and two additional parameters that adjust the reliance on historical trajectory data versus traditional road network segments. This setup ensures a flexible and effective routing solution.


To summarize, we make the following contributions:
\begin{itemize}
    \item We introduce \modelname{TrajRoute}, a new simple routing paradigm that bypasses the reliance on graph-based approaches and instead utilizes raw historical trajectory data to compute optimal routes. To the best of our knowledge, this is the first study to fully eliminate the current preprocessing pipeline.
    
    \item To address gaps in trajectory coverage, we integrate the road network into the same index as trajectories. We propose tunable parameters to control the balance between using historical trajectories and road network segments, ensuring flexible and complete route coverage while maintaining a unified approach.

    \item We experimentally verify the effectiveness of \modelname{TrajRoute} using real-world trajectory data. Our results show that the extracted routes are intuitive and comparable to those from an established routing engine in terms of distance and estimated travel time, while requiring lower maintenance costs.
\end{itemize}

The rest of the paper is organized as follows: Section 2 reviews related work regarding the utilization of trajectories in routing solutions. Section 3 formally defines the problem, and Section 4 describes our methodology. Section 5 discusses our experimental setup, dataset, and results. Section 6 concludes the paper.


\section{Related Work}

\textbf{Current routing approaches.} The vehicle routing problem has attracted significant attention over the years. Traditionally, this problem has been approached using graph-based models where the road network is represented by nodes (intersections or destinations) and edges (travel paths). Path-finding algorithms such as Bellman-Ford, Dijkstra, A*, and bidirectional A* are commonly employed to determine the shortest path between two nodes~\cite{pohl1971bi, hart1968formal}. To more effectively direct the search, algorithms like ALT~\cite{goldberg2005computing, goldberg2006reach}  have been developed. The ALT algorithm pre-selects a group of nodes called landmarks, and the shortest distances from every node in the network to these landmarks are calculated and stored in advance. During routing, a bidirectional A* search utilizes these pre-computed distances to estimate a lower bound distance between nodes for the heuristic computation. To accelerate computations, several pre-processing-based methods have also been proposed in the literature~\cite{abraham2011hub, chondrogiannis2016pardisp, geisberger2008contraction, sommer2014shortest, zhu2013shortest, demiryurek2011online}, such as Contraction Hierarchies (CH)~\cite{geisberger2008contraction}, which simplifies the graph by removing less important nodes, that once removed, would not affect the shortest path distances between the other, and more important, nodes. Our approach diverges from these traditional methods by not relying solely on graph-based representations of road networks.

\textbf{Trajectory-based routing approaches.} Conventional routing methods typically focus on finding the lowest cost paths based on distance or travel time. However, using historical trajectory data has opened up possibilities for more personalized or context-aware routing suggestions \cite{ceikute2015vehicle, guo2018learning, chen2011discovering, delling2015navigation, letchner2006trip, patel2006personalizing, yang2015toward, chang2011discovering, sacharidis2017finding, li2020spatial}. For example, Chang et al.~\cite{chang2011discovering} utilized historical trajectory segments to build a road network, recommending routes based on the most frequently traveled road segments. Ceikute et al.~\cite{ceikute2015vehicle} proposed a sophisticated scoring function of historical routes, considering both the frequency of traversals and the distinct number of drivers that took each route, but also temporal factors such as whether the travel occurred during peak traffic hours. Guo et al.~\cite{guo2018learning} developed a routing preference model that first clusters the vertices of the original road network into regions, creating a region graph. Then, routing preferences are learned from the available historical trajectories that connect some region pairs. These preferences are transferred to similar region pairs that are not connected by trajectories, ensuring that the approach will work even when trajectories are not available. Our approach differs from these methods in two distinct ways. First, unlike previous studies, we do not extract insights from the trajectory data and do not model scoring functions to rank trajectories based on these insights. Instead, we leverage the raw data directly to guide routing decisions. By directly following previous historical trips, we are implicitly recommending routes that might be more preferred by the drivers. Second, the aforementioned approaches still depend on a graph-based framework, while our approach is grid-based. Our innovations are designed to greatly reduce the maintenance cost of previous routing approaches.

\textbf{Travel time estimation.} Estimating the time of arrival (ETA) typically begins by identifying the fastest path in a network and aggregating the weights of its edges, which represent travel times. In static networks, these weights are constant. However, real-world travel times on road segments typically vary due to traffic congestion. To address this, historical trajectory data is leveraged to dynamically estimate average speeds on network edges. Because GPS data can be sparse and sometimes inaccurate, map-matching techniques are often employed to align the raw data on the corresponding road network~\cite{newson2009hidden}. Beyond these traditional methods, travel time estimation can also be performed directly from trajectories without explicitly computing the paths. Techniques range from simple regression models~\cite{ide2011trajectory} and decomposition methods~\cite{wang2014travel} to more recent deep learning approaches that model the spatial-temporal patterns embedded in trajectories to enhance the prediction accuracy of ETAs~\cite{wang2018learning, wang2018will, wu2019deepeta, fu2020compacteta, li2019learning, wang2022fine, yuan2020effective, lin2023origin}. While our approach, \modelname{TrajRoute}, primarily focuses on extracting the optimal path, it benefits from a substantial historical dataset that provides realistic travel time estimations as a secondary advantage.

To the best of our knowledge there is no previous work that directly utilized trajectories to extract routes from an arbitrary origin to a destination. One approach close to ours was presented in~\cite{anastasiou2019time}, where raw trajectory data were utilized for constructing isochrone and reverse isochrone maps. While our focused problem is different (single source to a single destination), this previous approach, though, fails when trajectory data does not fully cover that area.

\section{Preliminaries}

We cover the definitions of important concepts in our approach and introduce the problem.

\begin{definition}[Trajectory]
A \emph{trajectory} \( T \) is a sequence of GPS measurements, denoted as  $T = \langle ({p}_1, ts_1), ..., ({p}_{|T|}, ts_{|T|}) \rangle$, where $\text{p}_i = (lat_i, lon_i)$ represents the latitude and longitude coordinates of the $i_{th}$ measurement and $ts_i$ is the timestamp at which the measurement was recorded. $|T|$ denotes the length of the trajectory as the number of points. 
\end{definition}

\begin{definition}[Road Segment] A \emph{road segment} $S$ is a portion of a road network between two consecutive junctions or endpoints and can be represented as a polyline formed by a sequence of points $\langle p_1, p_2, ..., p_{|S|} \rangle$, where each ${p}_i = (lat_i, lon_i)$ represents the latitude and longitude coordinates of an intermediate point along the segment, $v_S$ denotes the legal travel speed limit and |S| is the length of the segment as the number of points. \end{definition}

\begin{definition}[Routing Path]
A \emph{routing path} \(P = \langle p_1, p_2, \ldots, p_{|P|} \rangle\) is a sequence of geographical points, where each $\text{p}_i = (lat_i, lon_i)$ represents the latitude and longitude coordinates of a point along the route, and $|P|$ is the total number of points that form the path. \(P\) can include points such that $p_i \in T$ for trajectory points or $p_i \in S$ for points along road segments.
\end{definition}

\begin{theorem*}
Let \( D \) be a database containing a collection of historical trajectories, and let \( R \) be the road network. Given a query tuple \( Q = (p_{\text{OR}}, p_{\text{DEST}}, \text{t}) \), where \( p_{\text{OR}} \) is the origin, \( p_{\text{DEST}} \) is the destination, and \( t \) is the departure time, the objective of \modelname{TrajRoute} is to compute a routing path \( P = \langle Q.p_{\text{OR}}, p_i, p_{i+1} \ldots, Q.p_{\text{DEST}} \rangle \) from \( Q.p_{\text{OR}} \) to \( Q.p_{\text{DEST}} \) by following trajectory data recorded at a time window $w_{time}$ close to \( Q.t \), i.e. $dt = Q.t \pm w_{time}.$ When trajectory data is unavailable for certain parts of the route, \modelname{TrajRoute} fills the gaps using road segments from \( R \), ensuring that a complete path is returned to the user.
\end{theorem*}

\section{The TrajRoute Approach}
In this section, we outline our approach for \modelname{TrajRoute}. Section~\ref{sec:index} introduces the index structure, followed by Section~\ref{sec:neighbors}, which discusses the identification of neighbors for the search algorithm. Section~\ref{sec:movement_cost} explains the computation of movement costs, while Section~\ref{sec:road_ntw} addresses the fallback to road network data when trajectory coverage is insufficient. Finally, Sec~\ref{sec:algo} describes the search algorithm used to compute the best routes in terms of travel time cost.

\subsection{Index Structure} \label{sec:index}

One of the main advantages of \modelname{TrajRoute} is its ability to operate directly on raw trajectory data. This is achieved through a grid-based index, which enables the quick extraction of trajectories near the start and destination locations of a user's query. To implement this, \modelname{TrajRoute} initializes an empty grid over the area of interest, where each cell has a uniform edge length of $n$ meters. Trajectories are then mapped to the grid cells based on the coordinates of their GPS measurements, with each measurement mapped to its corresponding cell. This grid-based approach not only structures the data in a way that supports efficient querying, but also accommodates minor GPS inaccuracies, ensuring that slightly deviating trajectories are still considered during query answering. As a result, the need for map matching the GPS points is eliminated.

Since traffic conditions vary greatly throughout the day, the relevance of certain trajectories for routing can change based on the time and date of travel. For example, during rush hours, drivers often choose longer but faster routes to avoid congestion, while at night, highways may be preferred due to lower traffic. To ensure that realistic route suggestions and accurate travel time estimates are returned to the user, the trajectories should be filtered according to the time and the day of the week they were recorded, matching with those of the user's queries. To achieve this, \modelname{TrajRoute} builds a B-tree on the timestamps of trajectories within each non-empty grid cell. During query processing, only the trajectories recorded that satisfy the temporal time window $w_{time}$ are considered. This allows \modelname{TrajRoute} to retrieve time-relevant trajectories without the need for creating, maintaining, and updating road segment traffic snapshots. 

\subsection{Identifying Neighbors} \label{sec:neighbors}

In traditional graph-based routing methods, neighbors are typically defined as nodes directly connected by an edge to the current node being explored. In TrajRoute, the grid structure offers an intuitive way to define neighbors. To that end, we consider two types of neighbors: the next GPS measurement in the current trajectory, which captures the natural, sequential movement along a continuous path, and nearby GPS measurements from other trajectories that are mapped to the same grid cell.

Formally, given a trajectory dataset \( D \), the index \( G \), and a query tuple \( Q = (p_{\text{OR}}, p_{\text{DEST}}, t) \), we define the neighbors $Nbr$ of a node~\footnote{In this context nodes are represented by points in our dataset.} \( p_k \) of trajectory \( T \in D \), as follows:

\begin{align}
{Nbr}(p_k) = & \{ p_{k+1} \mid k < |T| \} \: \cup \nonumber \\
    & \{ p_m \mid p_{m-1} \in \text{cell}(p_k, G), T' \neq T, |ts_{p_m} - ts_{p_k}| \leq w_{time} \}
\end{align}

\noindent Here, $p_{k+1}$ represents the next GPS measurement on $T$, if it exists. This captures the natural progression along a continuous path, and no temporal constraints are imposed on this transition. Furthermore, $p_m$ represents the $m_{th}$ measurement of another trajectory $T' \in D$ whose GPS measurement $p_{m-1}$ exists in the same cell as $p_k$ and was recorded at a close time and same date as $p_k$.

This formulation allows \modelname{TrajRoute} to continue exploring a trajectory by following the next point while also considering alternative routes by examining other trajectories whose GPS measurements are spatially nearby.

\subsection{Movement Cost} \label{sec:movement_cost}

Each neighbor is associated with a \textit{movement cost}, which influences its priority for processing and contributes to the overall cost of the final route. In \modelname{TrajRoute}, since the approach operates on raw trajectory data, these costs are naturally encoded in the dataset.

Let us assume that we are currently exploring node $p_k \in T$, and let us define \( p_m \) as one valid neighboring node. As mentioned before, we consider two types of neighbor transitions. The first type is moving along the same trajectory. In this case, we can move to the next GPS point of the same trajectory, \( p_m = p_{k+1} \) and the movement cost is based on their timestamp difference, that is \( |ts_{p_{k+1}} - ts_{p_k}| \).

The second type of neighbor transition is switching to a different trajectory \( T' \). For this transition to be valid, the previous point in \( T' \), \( p_{m-1} \), must lie in the same grid cell and within a temporal window \( w_{time} \) as \( p_k \). For this, a constant transition cost from \( T \) to \( T' \), called \( \tau_c \), is applied, which depends on the size of the grid cell. Intuitively, larger cells require higher transition costs, while for smaller cells, \( \tau_c \) can be as low as zero. After switching to \( T' \), the movement cost is computed as the time difference between \( p_m \) and \( p_{m-1} \), that is, \( \tau_c + |ts_{p_m} - ts_{p_{m-1}}| \).

Overall, we formally define formally the movement cost between two nodes $p_k$ and $p_m$ that belong to trajectories in $D$, $C_{traj}$ as:

\begin{equation}
C_\text{traj}(p_k, p_m) = 
\begin{cases}
    |ts_{p_{k+1}} - ts_{p_k}|  & \text{if } p_k, p_{m} \in T, m=k+1 \\
    \tau_c + |ts_{p_m} - ts_{p_{m-1}}| & \text{if } p_k \in T, p_m \in T'
\end{cases}
\end{equation}

\subsection{Fallback to Road Network} \label{sec:road_ntw}

\subsubsection{Road Network Integration}

\modelname{TrajRoute} finds the optimal path from an origin to a destination using the trajectories stored in $D$, assuming these trajectories adequately cover the geographical area of interest as well as satisfying the temporal constraints. However, there may be instances where the spatial or temporal coverage of the trajectories is insufficient. In such cases, the system should be able to switch between utilizing the available trajectories and the established road network to bridge gaps and ensure a complete path is returned to the user. This transition should occur without adding extra overhead to the pipeline.

To enable this, we index the road segments using the same grid structure, $G(n)$. Although the road network is typically represented as a graph, the roads themselves can also be expressed as linestrings and, therefore, they can be easily integrated into the index structure $G(n)$. However, unlike trajectories, road segments lack temporal information. As a result, the B-tree index is only used on trajectory timestamps and does not index the road segments. Algorithm~\ref{algo:index} outlines the overall indexing process.

\begin{algorithm}
\caption{Index Construction}
\begin{algorithmic}[1]
\Statex \textbf{INPUT:} A database of trajectories $D$, road network $R$, grid cell size $n$ (meters)
\Statex \textbf{OUTPUT:} Grid-based index $G$

\State Initialize grid $G$ with empty cells of size $n \times n$ covering the area of interest

\For{each trajectory $T_i \in D$}
    \State $pos_{I} \gets$ position of $T_i$ in $D$
    \For{each GPS measurement $m_j = (p_j, ts_j)$ in $T_i$}
        \State $pos_{J} \gets$ position of GPS measurement $m_j$ in $T_i$
        \State $cell\_id \gets$ FindCell($G$, $p_j$) 
        \State $B_{cell\_id} \gets$ B-tree associated with cell $cell\_id$ in grid $G$
        \State Insert $\langle ts_j, \langle pos_{I}, pos_{J} \rangle \rangle$ into $B_{cell\_id}$ 
    \EndFor
\EndFor


\For{each road segment $S_i \in R$}
    \State $pos_{I} \gets$ position of $S_i$ in $R$
    \For{each point $p_i$ in $S_i$}
        \State $pos_{J} \gets$ position of $p_j$ in $S_i$
        \State $cell\_id \gets$ FindCell($G$, $p_j$) 
        \State Insert $\langle pos_{I}, pos_{J} \rangle$ into $cell\_id$ 
    \EndFor
\EndFor

\State \textbf{return} $G$
\end{algorithmic}
\label{algo:index}
\end{algorithm}

\subsubsection{Adjusting Neighbors and Movement Costs}

With a unified indexing approach, extending the notion of neighbors to consider both trajectory and road points becomes straightforward. For road points, the next point on the polyline can be considered a natural neighbor, similar to traveling along road segments. Additionally, like trajectory neighbors, road points within the same grid cell can also be considered neighbors. 

The movement cost between road nodes is associated with the distance traveled and can be defined in two cases. First, for movement along the same road segment \( S \), the cost is determined by the distance between consecutive points on the polyline, divided by the speed \(v_S\) of \( S \). Second, for transitions to a different road segment \( S' \), the movement cost is a combination of a transition cost \( \tau_c \) and the distance-based movement cost on the new segment. 

Formally, the cost of movement $C_{road}$ between two nodes that belong to road segments can be expressed as:

\begin{align}
    C_\text{road}(p_k, p_m) = 
    \begin{cases} 
        \frac{\text{dist}(p_k, p_{k+1})}{v_S}, & \text{if } p_k, p_m \in S, m = {k+1} \\
        \tau_c + \frac{\text{dist}(p_{m-1}, p_m)}{v_{S'}}, & \text{if } p_k \in S, p_m \in S'
    \end{cases}
\end{align}

\noindent where \( \textit{dist} \) is the Haversine distance between the nodes.

The overall cost across neighbors can be consistently defined across both datasets: a point can progress along its path by following the next GPS measurement in a trajectory or the next point in a road segment while also considering nearby points from other trajectories or road segments within the same cell. 

The overall cost function $C_{base}$ that accounts for all the different types of transitions is defined as:

\begin{align} \label{eq:original_cost}
C_{base}(p_k, p_m) = 
\begin{cases} 
C_\text{traj}(p_k, p_m), & \text{if } p_m \in T \\
C_\text{road}(p_k, p_m), & \text{if } p_m \in S
\end{cases}
\end{align}

\subsubsection{Denoting Preference}
While our cost function computes travel time costs based on the neighbor type, it inherently favors road segments over trajectories. This is because the trajectories reflect real-time traffic and road conditions, and thus, they naturally exhibit higher costs. On the other hand, the road segments are associated with theoretical costs, which assume a constant average speed, underestimating the actual travel time, especially during peak traffic periods.

To account for this, we introduce a new parameter, $r_{\text{penalty}} \geq 0$, to penalize the cost of moving to road neighbors over trajectory neighbors. To that extent, we update the cost of movement $C_{adj}$ from a node $p_k$ to $p_m$ as follows:   

\begin{equation}
C_{adj}(p_{k}, p_{m}) = 
\begin{cases} 
(1 + r_{\text{penalty}}) \cdot C_{base}(p_{k}, p_{m}) & p_{m} \in S \\
C_{base}(p_{k}, p_{m}) & \text{otherwise}
\end{cases}
\end{equation}

While prioritizing the exploration of trajectory routes over road segments is effective, it is also more realistic and practical to prioritize continuous trips over trajectory segment transitions. This mirrors actual driving behaviors and helps preserve the temporal continuity of real-world trips. To achieve this, we introduce a second parameter, $rw$, which represents the \textit{continuity reward} for maintaining a continuous trip on the same historical trajectory. Now, final cost of movement $C$ between nodes $p_k$ and $p_m$ can be defined as:

\begin{equation} \label{eq:cost}
C(p_{k}, p_{m}) = 
\begin{cases} 
e^{-rw} \cdot C_{adj}(p_{k}, p_{m}) & p_{k}, p_{m} \in T, m=k+1 \\
C_{adj}(p_{k}, p_{m}) & \text{otherwise}
\end{cases} 
\end{equation}

Overall, the inclusion of these two parameters in the final cost function improves \modelname{TrajRoute}'s ability to generate routes that are not only cost-effective but also more closely aligned with typical driving patterns.

\begin{algorithm}
\caption{get\_neighbors}
\begin{algorithmic}[1]
\Statex \textbf{INPUT:} Grid Index $G$, current point $p_k$, trajectory dataset $D$, road network $R$, start point $p_{\text{OR}}$, end point $p_{\text{DEST}}$, continuity reward $rw$, road penalty factor $r_{penalty}$
\Statex \textbf{OUTPUT:} List of neighbors $L$ with their associated movement costs

\State Initialize list of neighbors $L \gets \emptyset$
\State $cell\_id \gets$ FindCell($G$, $p_k$) 
\State $P_{cell} \gets$ All points in $cell\_id$ from index $G$ 
\For{each $ptr = (pos_I, pos_J) \in P_{cell}$}
    \State Retrieve point $p_m$ using $ptr$ from $D$ or $R$
    \If{$p_k$ is a trajectory point}
        \If{$p_m$ is a trajectory point}
            \State $f_{p_m} \gets g(p_m)$
            \If{$p_k, p_m \in$ same trajectory}
                \State $f_{p_m} \gets e^{-rw} \times f_{p_m}$
            \EndIf
            \State $f_{p_m} \gets f_{p_m} + h(p_m)$
        \Else \Comment{$p_m$ is a road point}
            \State $f_{p_m} \gets (1+r_{penalty}) \times g(p_m) + h(p_m)$
        \EndIf
    \Else \Comment{$p_k$ is a road point}
        \If{$p_m$ is a trajectory point}
            \State $f_{p_m} \gets g(p_m) + h(p_m)$
        \Else \Comment{$p_m$ is a road point}
            \State $f_{p_m} \gets (1+r_{penalty}) \times g(p_m) + h(p_m)$
        \EndIf
    \EndIf
    \State $L \gets L \cup \langle p_m, f_{p_m} \rangle$
\EndFor

\State \textbf{return} $L$
\end{algorithmic}
\label{algo:get_neighbors}
\end{algorithm}

\begin{algorithm}
\caption{Path Finding in TrajRoute}
\begin{algorithmic}[1]
\Statex \textbf{INPUT:} Grid Index $G$, query tuple $Q$.

\Statex \textbf{OUTPUT:} Final path $P$, and $ETA$.

\State Initialize priority queue $PQ$, visited points set $visited$, and travel score set $g\_map$;
\If{$Q.p_{OR} == Q.p_{DEST}$} 
\State \Return \(\langle Q.p_{OR}, 0\rangle;\)
\EndIf

\State $f_0 \gets g(Q.p_{OR}) + h(Q.p_{OR})$;
\State $PQ \gets PQ \cup \langle Q.p_{OR}, f_0 \rangle$;
\While{$PQ$ not empty}
    \State $p_k \gets \text{node with the lowest cost from } PQ;$
    \State $PQ \gets PQ \setminus p_k$;
    \If{$dist(p_k, Q.p_{DEST}) < d_{thres}$}
        \State $Route, ETA \gets \text{Get final route associated with } p_k;$
        \State \Return $\langle Route, ETA \rangle;$
    \EndIf
    
    \State $visited \gets visited \cup p_k$;
    \State $neighbors \gets \text{ \textbf{call} get\_neighbors}(G, p_k);$
    \For{$\langle p_m, f_m \rangle \in neighbors$}
        \If{$p_m \notin visited$}
            \State $g_{p_m}\_{min} \gets p_m$ travel score from g\_map; 
            \If{$p_m \notin g\_map$ or $g(p_m) <  g_{p_m}\_{min}$}
                \State Update $g\_map$ with $g_{p_m}\_{min}$;
                \State $PQ \gets PQ \cup \langle p_{m}, f_m \rangle$;
            \EndIf
        \EndIf
    \EndFor
\EndWhile

\State \Return $\langle \emptyset, 0 \rangle$ \Comment{No path found}

\end{algorithmic} \label{algo:a_star}
\end{algorithm}

\subsection{Query Answering} \label{sec:algo}

We now detail the overall search algorithm, as presented in Algorithm~\ref{algo:a_star}. The algorithm initially receives a query tuple \(Q\), a grid index \(G\), and a vector \(vec\_data\) that contains all historical trajectories $D$ and road segments $R$. To illustrate the effectiveness of the approach, we describe how the A* search algorithm could be employed, given the new setup, to extract the optimal path $P$. We note that any pathfinding algorithm could be utilized to extract $P$, and that this choice is orthogonal to the approach.

A priority queue \(PQ\) is utilized to sort the neighbor points by their estimated travel costs in ascending order, starting with the query's start point \(Q.p_{\text{OR}}\), which is initially added to the queue.

During each iteration, the point \(p_k\) with the lowest estimated cost \(f_k\) is removed from \(PQ\) for expansion, with \(f\) defined as follows: 

\begin{align*}
f(p_{k}) = g(p_k) + h(p_{k}) \texttt{,} 
\end{align*}

\noindent where $g(p_k)$ is the travel cost from $Q.{p_{OR}}$ to $p_k$ and $h(p_k)$ is the heuristic estimate of travel cost from $p_k$ to $Q.{p_{DEST}}$. 

The travel cost \( g(p_k) \) is defined as the summation of the individual movement costs between consecutive nodes \( p_i \) along the current path $P$, starting from the query’s origin \( Q.p_{\text{OR}} \) to the current node \( p_k \). Formally, this can be expressed as:

\begin{equation}
g(p_k) = \sum_{i=1}^{|P|} \text{C}(p_{i-1}, p_i), \:\: p_i \in P
\end{equation}

For the heuristic estimate, we choose to implement a fast-to-evaluate heuristic, as typically used in open-source routing engines~\cite{valhalla}. To that extent, $h(p_{k})$ is defined as:

\begin{equation} \label{eq:heuristic}
    h(p_{k}) = \frac{dist(p_{k}, Q.p_{DEST})}{v_{max}} \texttt{,} 
\end{equation}

\noindent where $dist$ is the haversine distance and $v_{max}$ indicates the maximum allowable speed in the area of interest, i.e., the maximum speed limit in San Francisco is 70 mph. This heuristic always underestimates the actual travel cost of reaching $Q.p_{DEST}$, which makes it admissible. More tight heuristics could also be applied as well, such as the ones presented by~\citet{demiryurek2011online}. These heuristics rely on the precomputation of all the costs of all possible paths to form tighter heuristics and thus require heavy preprocessing. 

The neighbors of \(p_k\) are identified using the \texttt{get\_neighbors} function as outlined in Section~\ref{sec:neighbors}. This process involves determining the cell \(cell\_id\) that contains \(p_k\) and then examining each pointer \(ptr\) within cell \(cell\_id\). Each pointer leads to a trajectory or road segment, and the subsequent raw coordinate is considered as a potential neighbor node, denoted \(p_m\). The cost of transitioning from \(p_k\) to \(p_m\) is computed according to Equation~\eqref{eq:cost} and the heuristic cost $h$ is computed according to Equation~\eqref{eq:heuristic}. Since the cost of transitioning is adjusted to denote trajectory preference, the pre-adjusted cost of movement (Equation~\eqref{eq:original_cost}) is recorded to later help us with the computation of the Estimated Time of Arrival (ETA). Additionally, the sequence of coordinates visited is tracked to reconstruct the path once the destination is reached.

The \texttt{get\_neighbors} function returns a list of these neighbors. Each neighbor is added to \(PQ\) only if it has not been explored before or if the recalculated cost indicates a more optimal path than previously noted. This process repeats until the point removed from the PQ is within a predefined distance threshold $d_{thres}$ from the destination point \(Q.p_{\text{DEST}}\). The estimated time of arrival $ETA$ and the fastest path $ROUTE$ are returned to the user.

\section{Experimental Evaluation}
In this section, we present the evaluation of the proposed \modelname{TrajRoute} framework. We begin by detailing the dataset and the preprocessing steps involved to prepare it for the experiments, as well as the default hyperparameter values. Then, we outline the experimental setup, present and discuss the results of our experiments, demonstrating the effectiveness of \modelname{TrajRoute} in various scenarios.

\subsection{Experimental Setup}

\subsubsection{Dataset and Parameter Selection.}
\begin{figure}
    \centering
\includegraphics[width=0.75\linewidth]{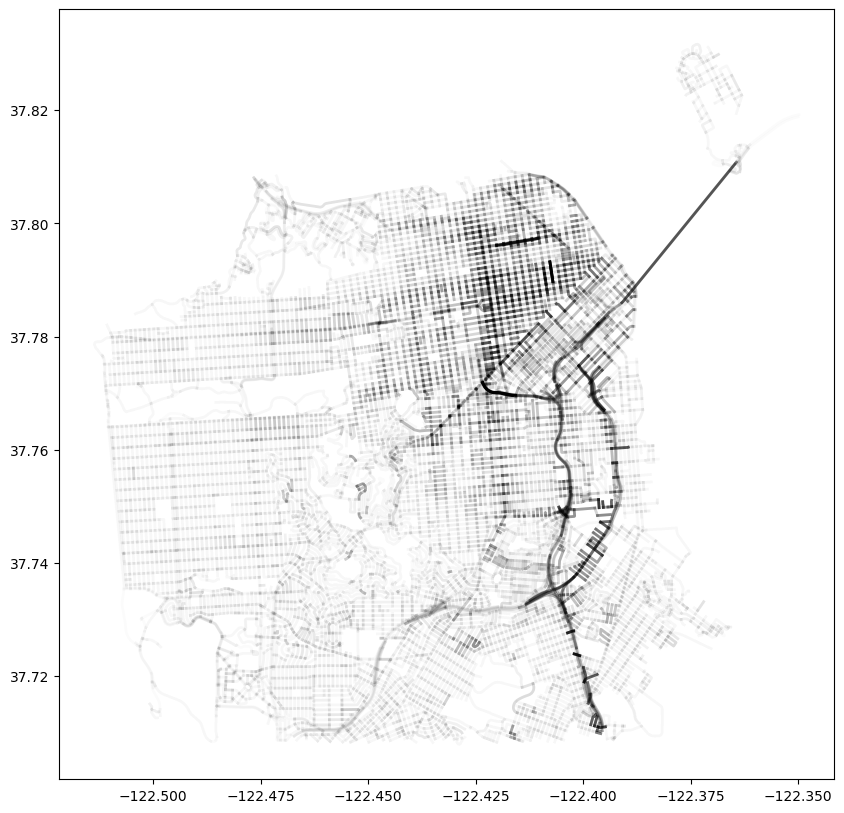}
    \caption{Density of spatial coverage for San Fransisco in our dataset. A darker color indicates a higher coverage for the segment.}
    \label{fig:spatial_cov}
    \Description{Density of spatial coverage for San Fransisco in our dataset. A darker color indicates a higher coverage for the segment.}
\end{figure}

We utilize a real-world taxi trajectory dataset collected in San Francisco, USA\footnote{\url{https://stamen.com/work/cabspotting/}}~\cite{comsnets09piorkowski}. The dataset contains over 11M records, corresponding to more than 1M trajectories. We remove trajectories that traveled less than 500 meters or less than 2 minutes during pre-processing. We used OpenStreetMap\footnote{\url{www.openstreetmap.org}} to extract the San Francisco area's road network and road segment travel speed limits. The dataset has a total of 27,279 roads. The trajectory dataset has a spatial coverage of 99.0\% of the extracted roads. This means that 99\% of the roads are traversed by at least one trajectory. Figure~\ref{fig:spatial_cov} illustrates the density of spatial coverage. Additionally, $24.83\%$ of the trajectories were collected during peak hours (7:00-10:00 AM and 4:00-7:00 PM), while $75.17\%$ occurred during off-peak hours. $62.51\%$ of the trajectories were collected on weekdays, with the remaining $37.49\%$ collected on weekends.

Regarding the parameter selection of \modelname{TrajRoute}, we set the grid size to $n = 100$ meters. Given that this is a relatively small size, we set the transition cost $\tau_c = 0$. Lastly, we look at trajectories that have occurred 30 mins before or after the query time ($w_{time}=30$) and set the threshold for identifying the destination point $d_{thres}$ to 100 meters.

\subsubsection{Hardware Configuration.}
All experiments were performed on an Ubuntu 18.04 server equipped with an AMD Ryzen 9 5900 12-Core Processor and 128GB of RAM. The processor has 12 cores (24 threads) with private L1 (32KB for data and 32KB for instructions) and L2 (512KB) caches and a shared L3 (64MB) cache.

\subsubsection{Evaluation and Metrics.}
For evaluation, we generate a set of origin-destination queries $\mathcal{Q}$ by randomly sampling start and end points from the trajectories in our dataset. The timestamp of the start point is used as the departure time. To evaluate the quality of the routes provided by \modelname{TrajRoute}, we compare the route length and travel time cost to those obtained from Azure Maps. Given that there are multiple route alternatives for each origin-destination pair, we believe that comparing travel time and route length provides more meaningful insights than directly comparing the similarity of the routes. For this comparison, we query the Azure Maps API~\footnote{\url{https://www.microsoft.com/en-us/maps/azure/azure-maps}} and set the departure time to match the one in our queries so that the results from Azure reflect accurate traffic conditions. For each comparison, we calculate the mean absolute error (MAE) for both the route length and travel time. The average route length for our query routes is 5.4 km, according to the results obtained by Azure Maps. The MAE is calculated as:

\[
MAE = \frac{1}{|\mathcal{Q}|} \sum_{i=1}^{n} |x_i - \hat{x}_i|
\]

\noindent where $x_i$ is the value obtained from \modelname{TrajRoute}, $\hat{x}_i$ is the value obtained from Azure Maps, and $|\mathcal{Q}|$ is the total number of queries.



\subsection{Quality of the generated paths}

In this section, we show that the quality of the routes generated by TrajRoute is comparable to or better than those produced by Azure Maps, which incorporates real-time and historical data and relies on a significant pre-processing and updating pipeline. We show this by analyzing the impact of TrajRoute’s tunable parameters, the road penalty $r_{penalty}$, and the continuity reward ($rw$) on route quality. By varying these parameters, we generate different routes and compare them against those provided by Azure Maps to assess their quality.

\begin{figure}[ht]
    \centering
    \begin{subfigure}[b]{0.5\columnwidth}
        \centering
        \includegraphics[width=\linewidth]{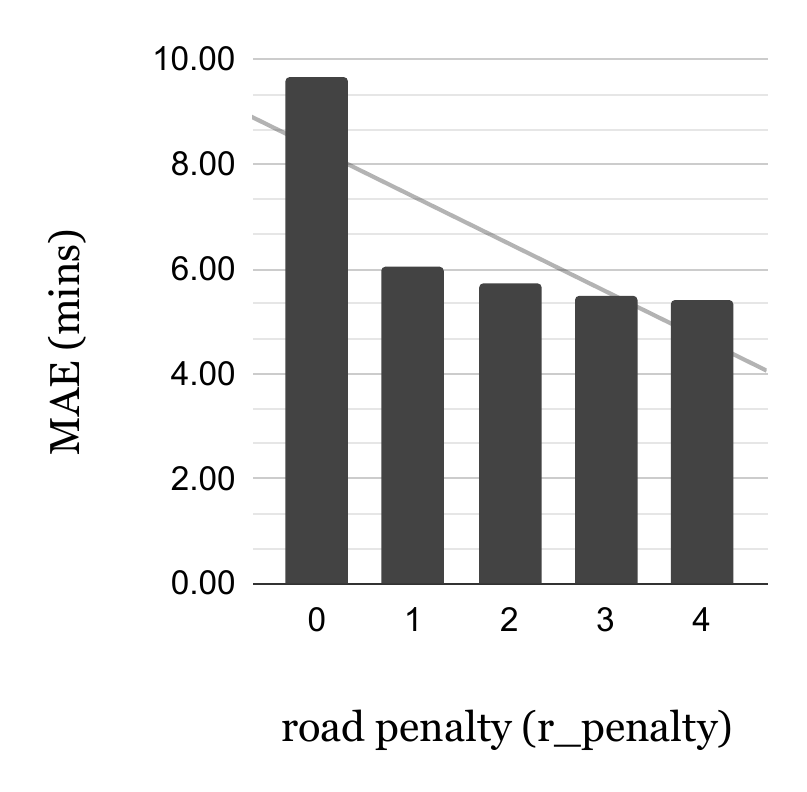}
        \caption{MAE of route travel time}
        \label{fig:road_penalty_time}
    \end{subfigure}%
    \begin{subfigure}[b]{0.5\columnwidth}
        \centering
        \includegraphics[width=\linewidth]{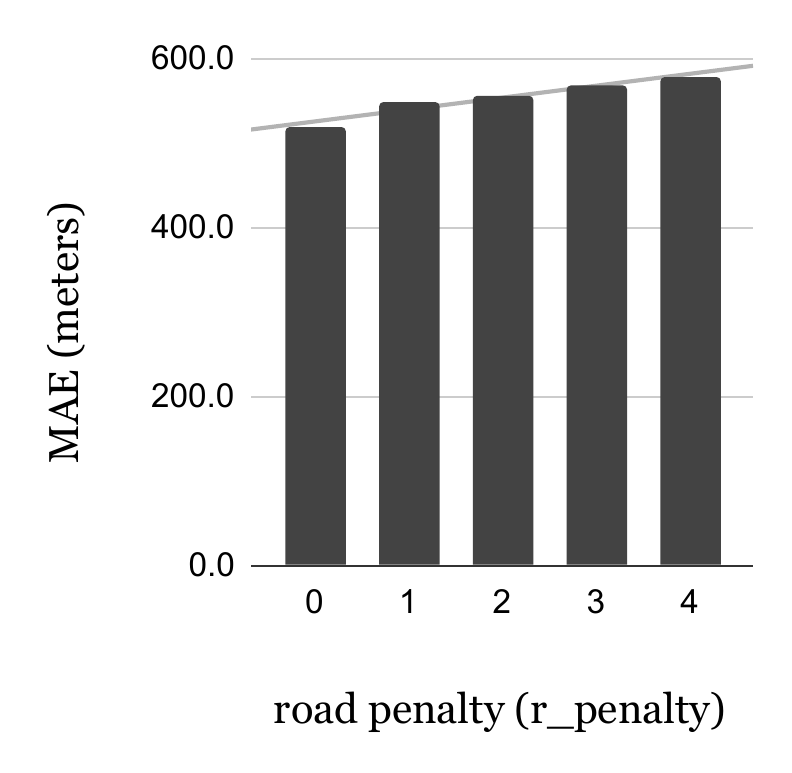}
        \caption{MAE of route distance}
        \label{fig:road_penalty_distance}
    \end{subfigure}
    \caption{Results for different $r_{\text{penalty}}$ values $\in$ [0, 4].}
    \label{fig:road_penalty}
    \Description{Results for different $r_{\text{penalty}}$ values $\in$ [0, 4].}
\end{figure}

We begin by focusing on the effect of the road penalty, $r_{penalty}$, by testing values in the range $[0, 4]$. This range reflects the trade-off between real-world travel conditions, as captured by trajectories, and the idealized conditions reflected in the road network. Figure~\ref{fig:road_penalty} shows the results for different values of $r_{penalty}$. As the road penalty increases — discouraging the use of road segments and encouraging greater reliance on historical trajectories — we observe a decreasing trend in the mean absolute error for travel time estimation (i.e., the estimated time for drivers to reach their destinations) in Figure~\ref{fig:road_penalty_time}. This indicates that relying more on historical trajectories allows \modelname{TrajRoute} to better account for real-world traffic patterns, leading to more accurate travel time predictions. These predictions align more closely with those from Azure Maps, which incorporates both real-time and historical traffic data into its calculations.
We also notice that improvements diminish when $r_{penalty} > 3$. This is likely because 75\% of the trajectories were collected during off-peak times, meaning that for this portion of the dataset, \modelname{TrajRoute} can effectively prioritize trajectories without requiring to excessively penalize the use of road segments.

Additionally, we report the distance differences between the routes generated by \modelname{TrajRoute} and Azure Maps to demonstrate that the routes generated by our approach are reasonable (i.e., similar in length). In this case, we observe a slight increase in path distance as $r_{penalty}$ increases, as shown in Figure~\ref{fig:road_penalty_distance}. This can be explained by the fact that as \modelname{TrajRoute} relies more heavily on trajectories, it may select routes that are slightly longer but preferred by drivers. These small deviations (less than 600 meters) might increase the total trip distance but reflect more realistic driver choices.

\begin{figure}[ht]
    \centering
    \begin{subfigure}[b]{0.45\columnwidth}
        \centering
        \includegraphics[width=\linewidth]{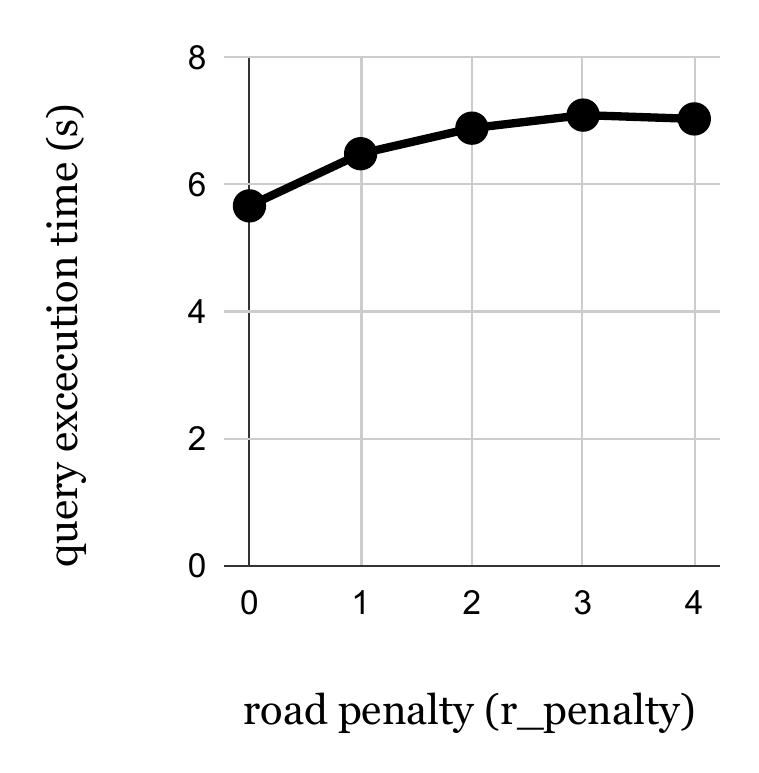}
        \caption{Impact of $r_{penalty}$ on \\ avg. query execution time.}
        \label{fig:suba_avgt}
    \end{subfigure}%
    \begin{subfigure}[b]{0.45\columnwidth}
        \centering
        \includegraphics[width=\linewidth]{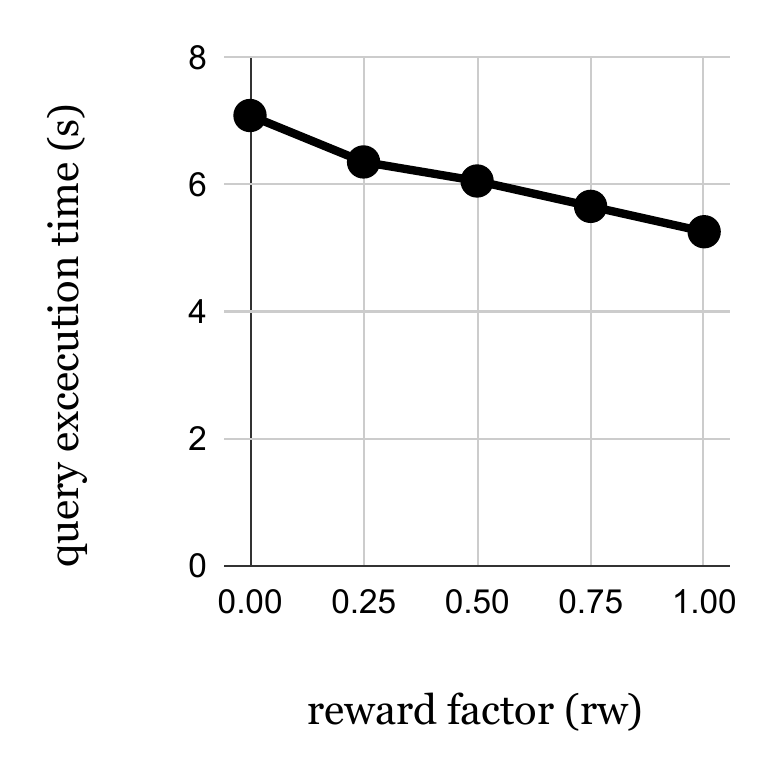}
         \caption{Impact of $rw$ on \\ avg. query execution time.}
         \label{fig:subb_avgt}
    \end{subfigure}
    \caption{Average query execution time.}
    \label{fig:avg_q}
    \Description{Average query execution time.}
\end{figure}

We also look at the impact of $r_{penalty}$ on query execution time. As shown in Figure~\ref{fig:suba_avgt}, as the penalty increases, there is a gradual rise in average execution time. This increase is expected, as higher penalties force \modelname{TrajRoute} to rely more heavily on trajectory data, resulting in a larger search space and more paths to evaluate. However, the increase in execution time is relatively small, suggesting that \modelname{TrajRoute} can manage the shift from road segments to trajectories without introducing significant delays (less than 1 second with our modest hardware).

\begin{figure}[ht]
    \centering
    \begin{subfigure}[b]{0.5\columnwidth}
        \centering
        \includegraphics[width=\linewidth]{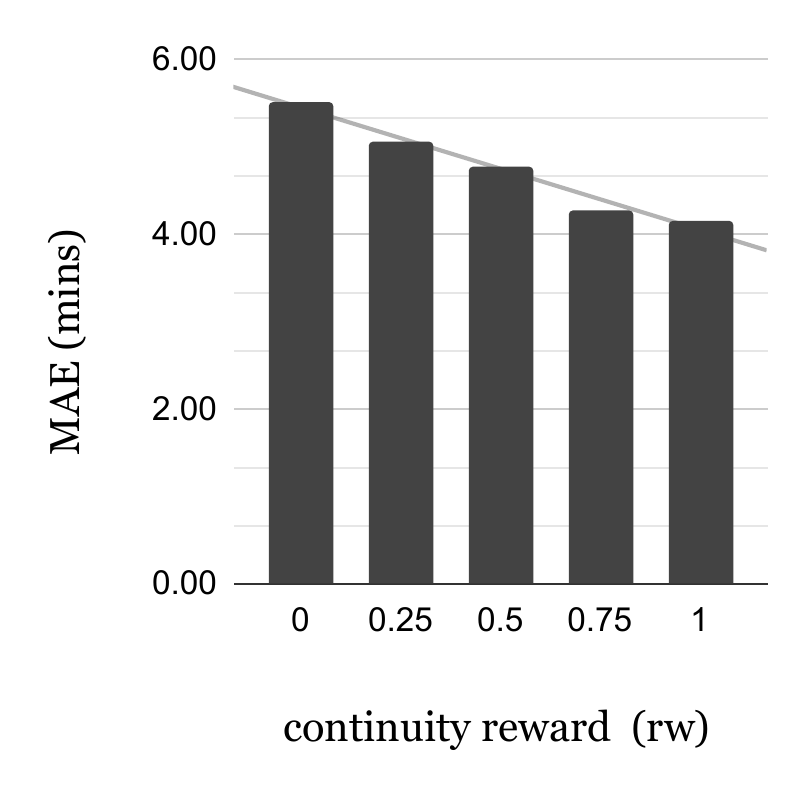}
        \caption{MAE of route travel time}
        \label{fig:cont_rw_time}
    \end{subfigure}%
    \begin{subfigure}[b]{0.5\columnwidth}
        \centering
        \includegraphics[width=\linewidth]{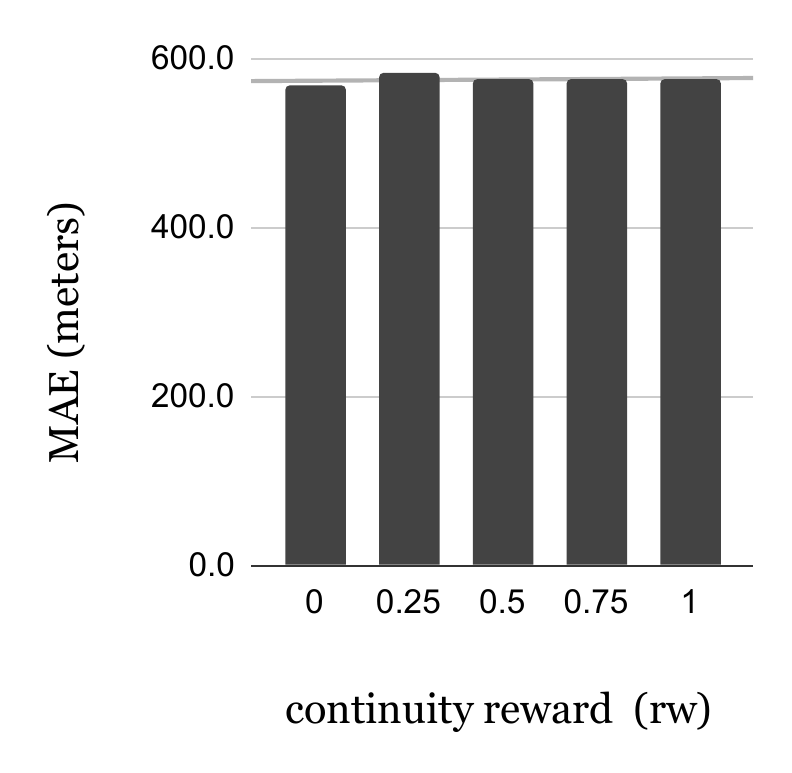}
        \caption{MAE of route distance}
        \label{fig:cont_rw_distance}
    \end{subfigure}
    \caption{Results for different $rw$ values $\in$ [0, 1].}
    \label{fig:cont_rw}
    \Description{Results for different $rw$ values $\in$ [0, 1].}
\end{figure}

Next, we examine the effect of the continuity reward ($rw$), which encourages following the same trajectory over switching between different trajectories or road segments. For this experiment, we set $r_{penalty}=3$. As shown in Figure~\ref{fig:cont_rw_time}, we again observe a decrease in the MAE for time travel cost as $rw$ increases. This suggests that following the same trajectory provides more accurate time estimates. This observation is intuitive, as sticking to a single trajectory closely mimics real-world driver behavior and allows \modelname{TrajRoute} to capture the real-time conditions of a specific trip. By doing so, \modelname{TrajRoute} is able to implicitly encode how a driver navigates and adjusts to conditions along the route, leading to more accurate travel time calculations that align even closer with those of Azure Maps, which computes routes that are sensitive to estimated traffic speeds.

At the same time, we observe in Figure~\ref{fig:cont_rw_distance} that the MAE for path distance remains relatively stable across all values of $rw$. This stability suggests that while the reward encourages staying on the same trajectory, it does not cause significant deviations in the overall route distance. The small values of $rw$ ensure that the approach avoids blindly following trajectories without considering optimal paths, maintaining reasonable route distances. 

Regarding performance for the $rw$, Figure~\ref{fig:subb_avgt} shows a steady decrease in average query execution time as the reward increases. Since continuity rewards encourage the system to stick to the same trajectory rather than switching between multiple trajectories or road segments, the search space is reduced, and fewer paths need to be explored, leading to faster query execution times.

\subsection{Varying Spatial Coverage}
As mentioned earlier, \modelname{TrajRoute} performs best when the trajectory dataset provides sufficient coverage of the area of interest. To assess how the framework behaves when trajectory coverage is limited and to explore its reliance on the road network as a fallback mechanism, we create trajectory sets with varying levels of coverage: 5\%, 25\%, 50\%, 75\%, and 100\% of the area. The coverage sets are built by incrementally adding trajectories to cover a specific percentage of the road network. Once the target coverage is reached, we continue to include trajectories that fully overlap with the already-covered roads, ensuring that all the trajectories covering that percentage of the area are added. For this experiment, we set $r_{penalty}=3$ and $rw=0.75$.

\begin{figure}[ht]
    \centering
    \begin{subfigure}[b]{0.5\columnwidth}
        \centering
        \includegraphics[width=\linewidth]{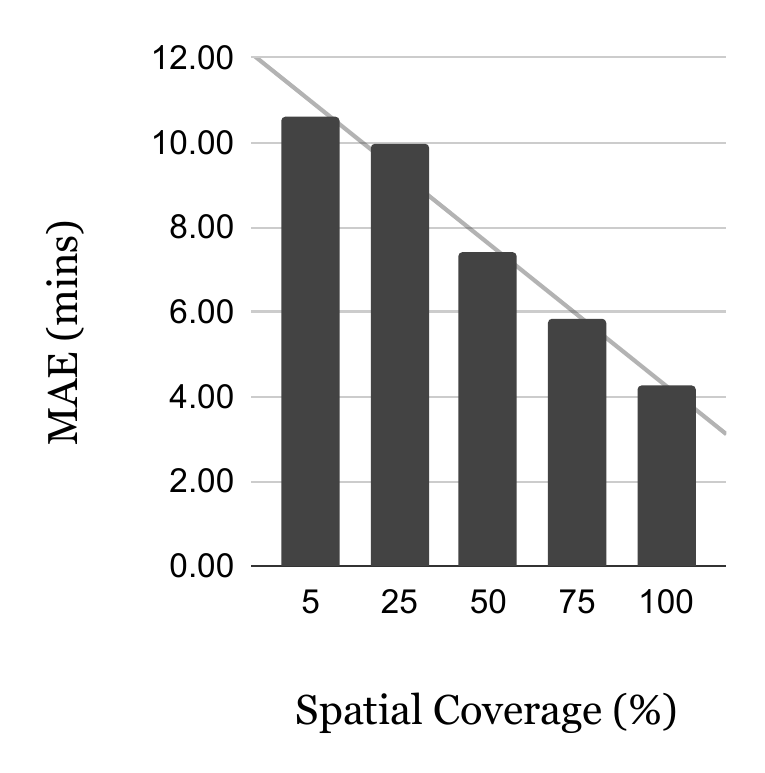}
        \caption{MAE of route travel time}
        \label{fig:spcov_a_time}
    \end{subfigure}%
    \begin{subfigure}[b]{0.5\columnwidth}
        \centering
        \includegraphics[width=\linewidth]{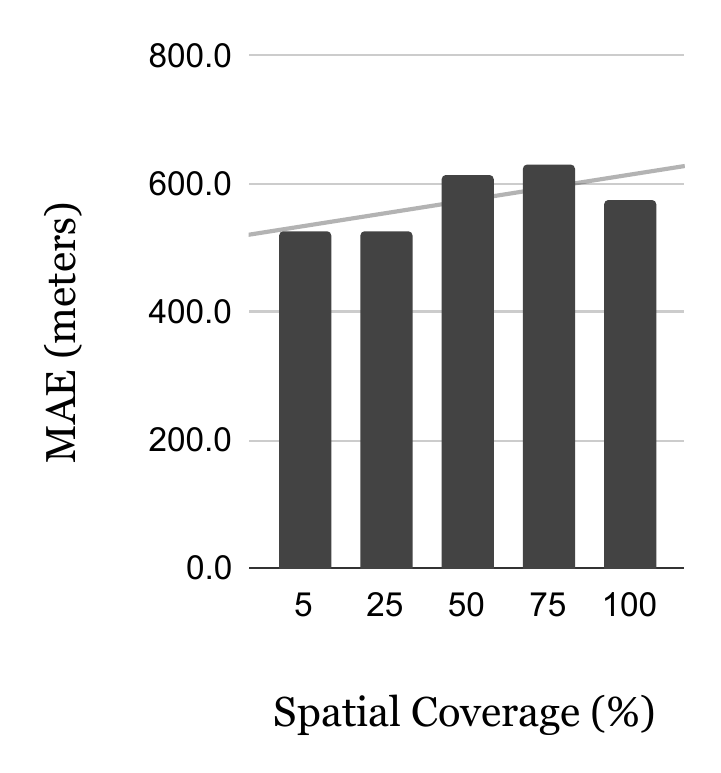}
        \caption{MAE of route distance}
        \label{fig:spcov_a_distance}
    \end{subfigure}
    \caption{Results for different levels of spatial coverage.}
    \label{fig:spcov_a}
    \Description{Results for different levels of spatial coverage.}
\end{figure}

\begin{figure}[ht]
    \centering
    \begin{subfigure}[b]{0.5\columnwidth}
        \centering
        \includegraphics[width=\linewidth]{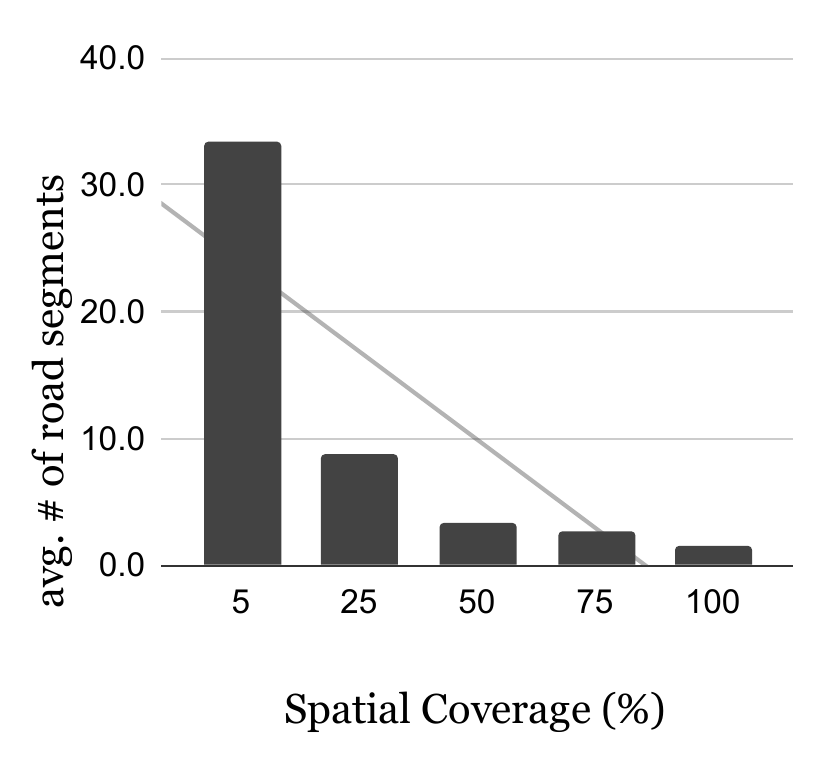}
    \end{subfigure}%
    \caption{Average number of road segments per route for different spatial coverage levels.}
    \label{fig:avg_roads}
    \Description{Average number of road segments per route for different spatial coverage levels.}
\end{figure}

Figure~\ref{fig:spcov_a}  shows the results for different coverage levels, presenting the MAE for both travel time and route length compared to Azure Maps. As the spatial coverage increases, we observe a clear improvement in the travel time estimates in Figure~\ref{fig:spcov_a_time}, with the MAE for travel time decreasing significantly. This demonstrates that as trajectory coverage increases, TrajRoute relies less on the road network’s predefined assumptions and more on actual historical trajectories, allowing it to better capture real-world driving conditions. In cases where certain road segments are not covered by any trajectories, TrajRoute must assume a fixed travel time based on the speed limit of those segments, which can degrade the accuracy of its ETA calculations. However, with more trajectory data, TrajRoute is able to more accurately reflect real travel conditions, resulting in a closer alignment with the travel time estimates provided by Azure Maps.

\begin{figure*}[ht]
    \centering
    \begin{subfigure}[b]{0.33\linewidth}
        \centering
        \includegraphics[width=0.8\linewidth]{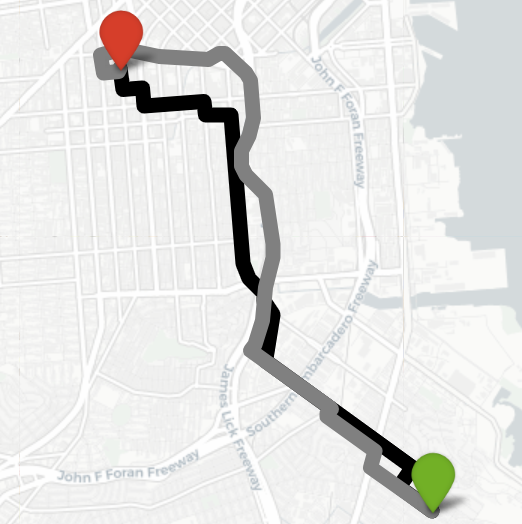}
        \caption{Route for $r_{penalty}=0$ and $rw=0$. \\
        \modelname{TrajRoute} ETA: 10 mins.
        }
        \label{fig:ex1_sub1}
    \end{subfigure}%
    \begin{subfigure}[b]{0.33\linewidth}
        \centering
        \includegraphics[width=0.85\linewidth]{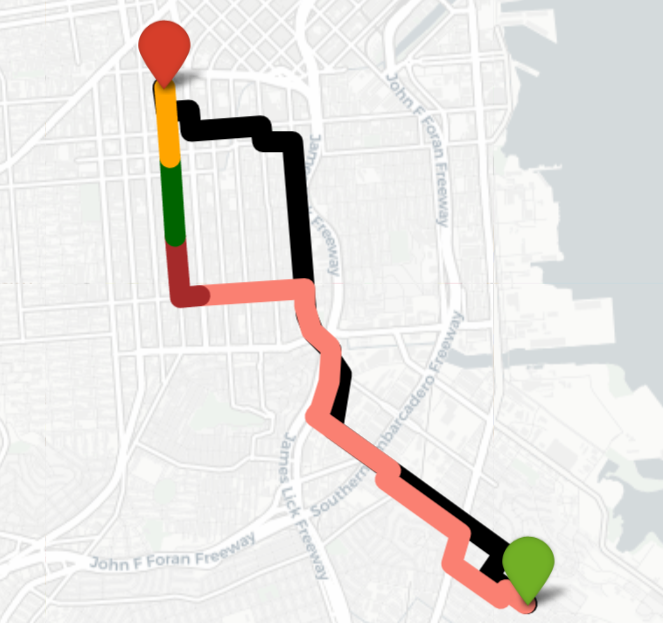}
        \caption{Route for $r_{penalty}=3$ and $rw=0$.\\
        \modelname{TrajRoute} ETA: 16.21 mins.
        }
        \label{fig:ex1_sub2}
    \end{subfigure}%
    \begin{subfigure}[b]{0.33\linewidth}
        \centering
        \includegraphics[width=0.87\linewidth]{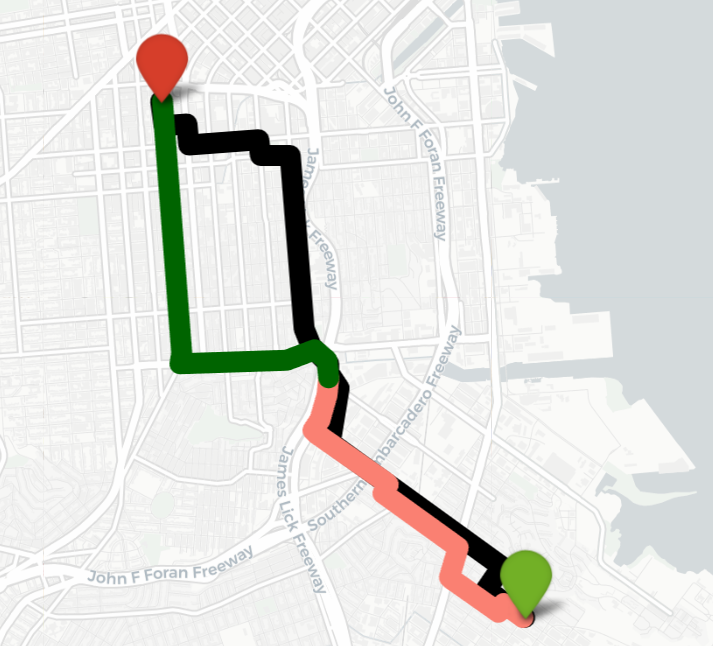}
        \caption{Route for $r_{penalty}=3$ and $rw=0.75$. \\
        \modelname{TrajRoute} ETA: 19.53 mins.
        }
        \label{fig:ex1_sub3}
    \end{subfigure}
    \caption{Example of routes computed by \modelname{TrajRoute} for different $r_{penalty}$ and $rw$ values. The departure time for this query is 06:01 PM, and Azure Maps ETA is 21 mins. Grey color represents road segments, black color represents the route generated by Azure Maps. Other colors represent different trajectories. }
    \label{fig:example1}
    \Description{Example of routes computed by TrajRoute for different $r_{penalty}$ and $rw$ values. The departure time for this query is 06:01 PM, and Azure Maps ETA is 21 mins. Grey color represents road segments, black color represents the route generated by Azure Maps. Other colors represent different trajectories.}
\end{figure*}

\begin{figure*}[ht]
    \centering
    \begin{subfigure}[b]{0.33\linewidth}
        \centering
        \includegraphics[width=0.85\linewidth]{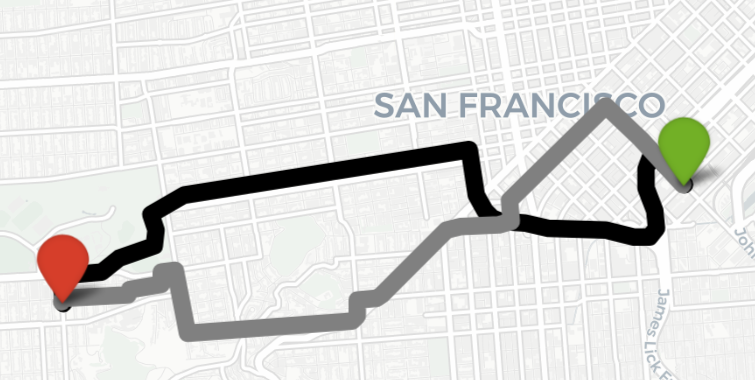}
        \caption{Route for $r_{penalty}=0$ and $rw=0$.\\ \modelname{TrajRoute} ETA: 8.3 mins.}
        \label{fig:ex2_sub1}
    \end{subfigure}%
    \begin{subfigure}[b]{0.33\linewidth}
        \centering
        \includegraphics[width=0.85\linewidth]{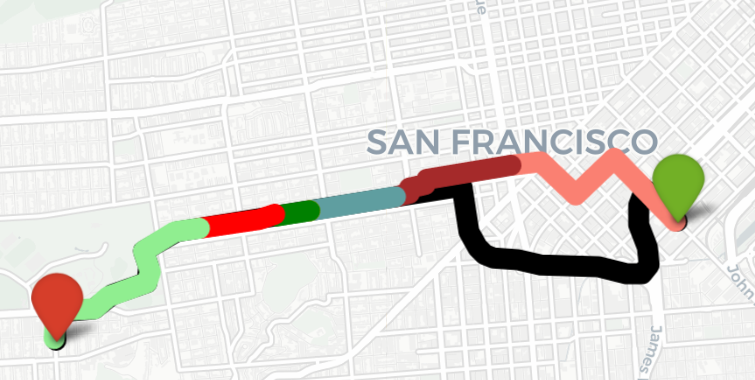}
        \caption{Route for $r_{penalty}=3$ and $rw=0$. \\ \modelname{TrajRoute} ETA: 10.15 mins.}
        \label{fig:ex2_sub2}
    \end{subfigure}%
    \begin{subfigure}[b]{0.33\linewidth}
        \centering
        \includegraphics[width=0.85\linewidth]{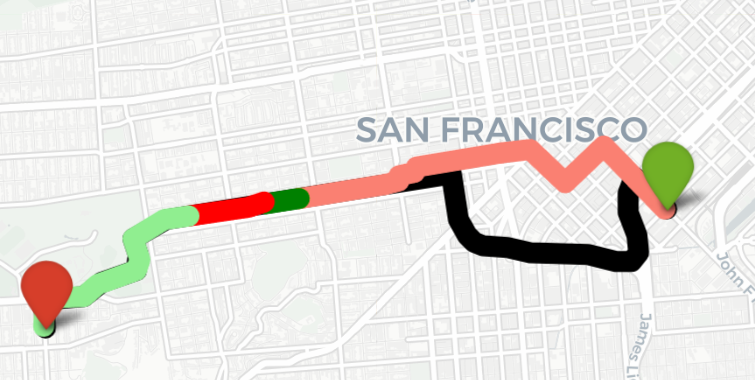}
        \caption{Route for $r_{penalty}=3$ and $rw=0.75$. \\
        \modelname{TrajRoute} ETA: 11.58 mins.
        }
        \label{fig:ex2_sub3}
    \end{subfigure}
    \caption{Example of routes computed by \modelname{TrajRoute} for different $r_{penalty}$ and $rw$ values. The departure time for this query is 01:25 AM, and Azure Maps ETA is 12 mins. Grey color represents road segments, black color represents the route generated by Azure Maps. Other colors represent different trajectories.}
    \label{fig:example2}
    \Description{Example of routes computed by TrajRoute for different $r_{penalty}$ and $rw$ values. The departure time for this query is 01:25 AM, and Azure Maps ETA is 12 mins. Grey color represents road segments, black color represents the route generated by Azure Maps. Other colors represent different trajectories.}
\end{figure*}

In contrast, in Figure~\ref{fig:spcov_a_distance}, the MAE for route distance does not follow a clear pattern, with some fluctuations observed at 50\% and 75\% coverage. These fluctuations can be explained by TrajRoute's reliance on the available trajectory data, which, at intermediate coverage levels, may not always offer the most direct or optimal routes. As a result, the system may select paths that slightly deviate in distance compared to the ones returned by Azure Maps. It’s important to note that even systems like Azure Maps and other standard navigation platforms, such as Google Maps, which rely on trajectory data to infer traffic and travel times for road segments, would encounter similar issues. If these systems had limited trajectory data for certain road segments, they too would have to rely on predefined speed limits, resulting in similar degradation in travel time accuracy. Overall, the distance error remains relatively stable across all coverage levels, indicating that the system can maintain accurate path lengths, whether relying on the road network or incomplete trajectory data. The heavy reliance of \modelname{TrajRoute} on trajectory data is further supported by Figure~\ref{fig:avg_roads}, which shows the average number of road segments used to form routes at each coverage level. We observe that the number of road segments used drops dramatically at the 25\% coverage level, indicating that \modelname{TrajRoute} prefers trajectories over roads whenever they are available.

\subsection{Increasing the Temporal Threshold}

\begin{figure}[ht]
    \centering
    \begin{subfigure}[b]{0.5\columnwidth}
        \centering
        \includegraphics[width=\linewidth]{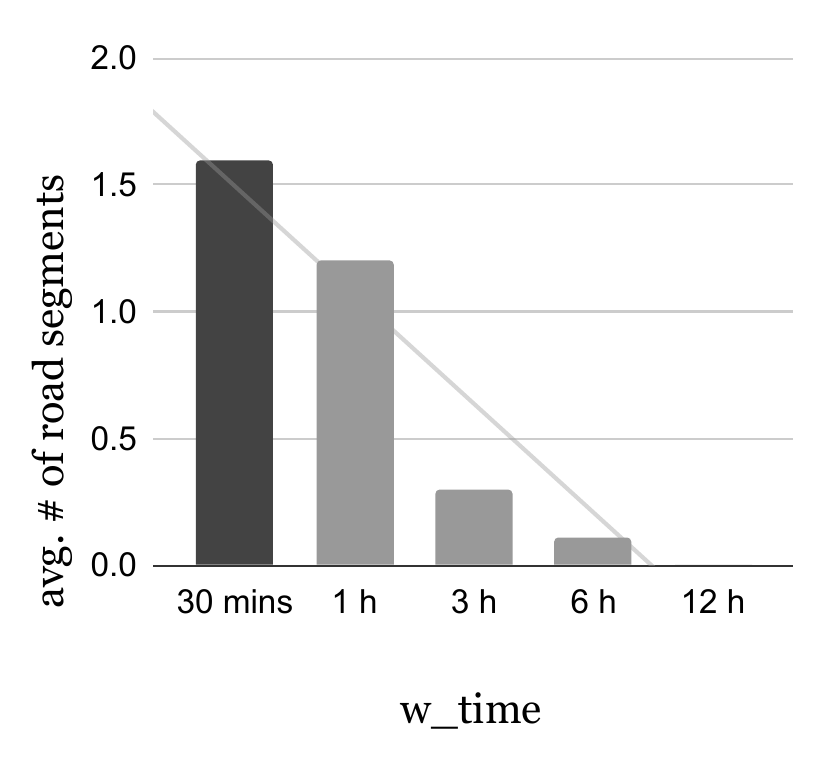}
        \caption{Avg. \# of road segments}
    \end{subfigure}%
    \begin{subfigure}[b]{0.5\columnwidth}
        \centering
        \includegraphics[width=\linewidth]{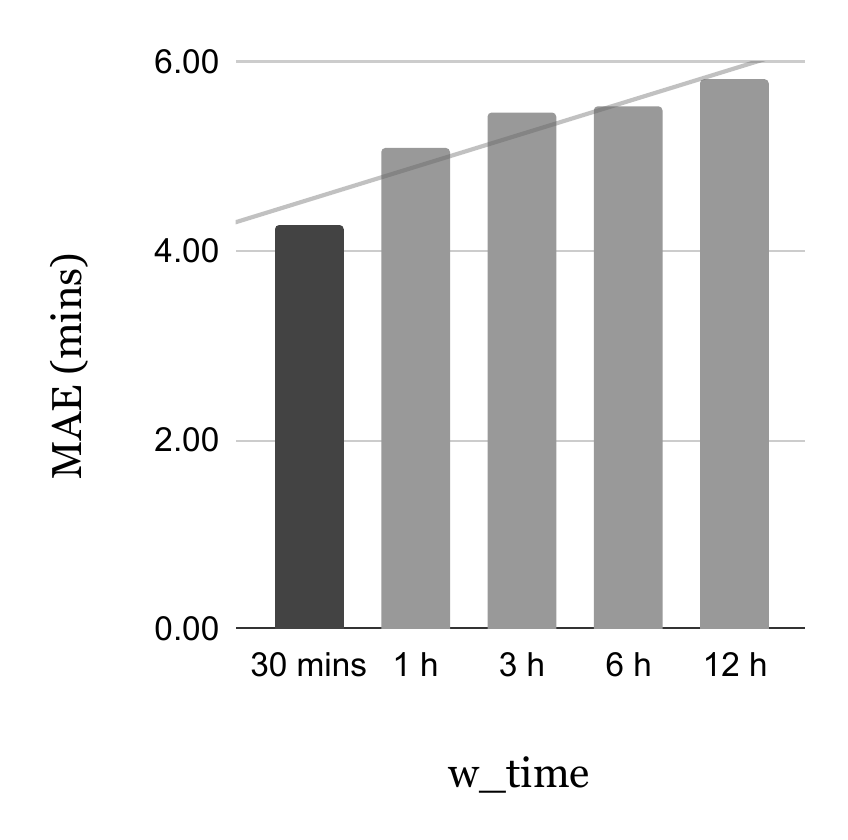}
        \caption{MAE of route travel time}
    \end{subfigure}
    \caption{Results for different levels of temporal coverage.}
    \label{fig:tcov}
    \Description{Results for different levels of temporal coverage.}
\end{figure}

In Figure~\ref{fig:avg_roads}, we observe that even with 100\% spatial coverage, there are still cases where road segments are used to form the final routes. We attribute this to the lack of available trajectories at all times. To demonstrate this, we increase the temporal window ($w_{\text{time}}$), which was initially set to 30 minutes, to values from 1 hour to 12 hours. 

Figure~\ref{fig:tcov} presents our results. As the temporal window increases, \modelname{TrajRoute} uses fewer road segments, and by the time the window reaches 12 hours, road segments are no longer used at all. This shows that a larger temporal window provides more trajectories to choose from, reducing the need to rely on the road network. However, as the temporal threshold widens, the MAE of estimated travel time, compared to Azure Maps, also increases. This is reasonable because using trajectories that were recorded further away in time from the queries may not reflect the current traffic conditions, leading to less accurate ETA predictions. Despite this, the increase in ETA remains within acceptable bounds, suggesting that even with relaxed temporal constraints, TrajRoute can provide reasonable travel time estimates, though with reduced real-time accuracy. We remind the reader that with adequate temporal coverage—such as the large, almost complete spatial and temporal trajectory data available to all standard navigation systems—there would be no need to relax the temporal threshold.

\subsection{Visual Evaluation}

In this section, we present real-world examples of the routes computed by \modelname{TrajRoute} for different values of road penalty, $r_{penalty}$, and continuity reward, $rw$. 

When both $r_{penalty}$ and $rw$ are set to 0, as shown in Figures~\ref{fig:ex1_sub1} and~\ref{fig:ex2_sub1}, \modelname{TrajRoute} primarily generates paths using road segments. As explained earlier, this is expected because road segments are associated with theoretical costs, assuming a constant average speed, which often underestimates actual travel time. Additionally, we observe that the generated routes tend to favor highways, where higher speed limits contribute to faster-estimated travel times. As a result, the ETA estimates from \modelname{TrajRoute} differ more from those of Azure Maps. We reiterate that our comparison assumes that Azure Maps has access to up-to-date traffic data, which makes this comparison somewhat unfair. If Azure Maps lacked trajectory data for those road segments, it would also default to predefined speed limits, potentially generating the same ETA and path as \modelname{TrajRoute}.

When $r_{penalty}$ is increased to 3, as shown in Figures~\ref{fig:ex1_sub2} and~\ref{fig:ex2_sub2}, \modelname{TrajRoute} relies more heavily on trajectories. This shift improves the accuracy of the ETA estimates, as trajectories better capture real-world traffic patterns. Even though the example in Figure~\ref{fig:example2} represents a scenario without traffic, relying more on trajectory data allows \modelname{TrajRoute} to account for real-world driving factors, such as traffic light stops, intersection delays, and speed variations. These factors are inherently captured in the trajectory data because they reflect actual driver behavior and conditions encountered on the road, such as waiting at stoplights or slowing down for turns. As a result, \modelname{TrajRoute} provides more accurate ETA calculations, even in scenarios where traffic is minimal, by incorporating these subtle time costs that are not accounted for in static road network data. Additionally, we observe that the system may use a variety of different trajectories to form the final routes. As demonstrated in Figure~\ref{fig:ex1_sub3}, by setting $rw$ to 0.75, \modelname{TrajRoute} encourages continuity by favoring single trajectories, reducing the frequency of switching between different trajectories. This further improves ETA estimates, bringing them even closer to those of Azure Maps, as the system more closely follows real-world driving behavior.

These examples confirm the observations made in the previous sections, illustrating how adjusting $r_{penalty}$ and $rw$ allows \modelname{TrajRoute} to better align with real-world conditions and improve the accuracy of its routes and ETA calculations.


\section{Conclusion}

In this paper, we introduced \modelname{TrajRoute}, a novel trajectory-based routing approach that eliminates the costly preprocessing steps employed by current navigation pipelines by directly leveraging raw trajectory data to compute efficient routes. Our experiments show that, with sufficient and up-to-date trajectory data, \modelname{TrajRoute} produces realistic routes with travel times and distances closely matching those from Azure Maps. This demonstrates \modelname{TrajRoute}'s ability to implicitly capture real-world factors such as traffic patterns, intersections, traffic lights, and stop costs, all while requiring low maintenance costs. Without up-to-date and sufficient trajectory data, both our approach, TrajRoute, and standard navigation systems like Azure Maps and Google Maps, which rely on trajectory data to infer traffic and travel times for road segments, would experience degraded performance. In the future, we plan to deploy \modelname{TrajRoute} within an end-to-end navigation pipeline.  We acknowledge that for the final step of a real pipeline, we need to map-match TrajRoute’s recommended path. However, we would like to emphasize that this map-matching of the final results is significantly less computationally intensive compared to map-matching all incoming collected trajectories at the start of the pipeline and as new trajectory data becomes available.


\bibliographystyle{ACM-Reference-Format}
\bibliography{references}

\end{document}